\documentclass[prd, aps, superscriptaddress, preprintnumbers, twocolumn, floatfix, nofootinbib]{revtex4-1}

\usepackage{amsfonts}
\usepackage{amsmath}
\usepackage{amssymb}
\usepackage{bm}
\usepackage{dcolumn}
\usepackage{graphicx}
\usepackage[latin1]{inputenc}
\usepackage{latexsym}
\usepackage{rotating}
\usepackage{graphicx}
\usepackage{color}
\usepackage{float}
\usepackage{epsfig}
\usepackage{soul}

\usepackage[unicode=true,pdfusetitle,
 bookmarks=true,bookmarksnumbered=false,bookmarksopen=false,
 breaklinks=false,pdfborder={0 0 1},backref=false,colorlinks=true]
 {hyperref}
\hypersetup{
 linkcolor=blue, citecolor=magenta, urlcolor=red, filecolor=blue}

\begin{document}

\title {T-dual cosmological solutions in double field theory }

\author{Robert Brandenberger}
\email{rhb@physics.mcgill.ca}
\affiliation{Physics Department, McGill University, Montreal, QC, H3A 2T8, Canada}

\author{Renato Costa}
\email{Renato.Santos@uct.ac.za}
\affiliation{Cosmology and Gravity Group, Dept. of Mathematics and Applied Mathematics,
University of Cape Town, Rondebosch 7700, South Africa}
\affiliation{Physics Department, McGill University, Montreal, QC, H3A 2T8, Canada}

\author{Guilherme Franzmann}
\email{guilherme.franzmann@mail.mcgill.ca}
\affiliation{Physics Department, McGill University, Montreal, QC, H3A 2T8, Canada}

\author{Amanda Weltman}
\email{amanda.weltman@uct.ac.za}
\affiliation{Cosmology and Gravity Group, Dept. of Mathematics and Applied Mathematics,
University of Cape Town, Rondebosch 7700, South Africa}

\date{\today}

\begin{abstract}

We investigate the cosmological solutions coming from the double field theory equations of motion after coupling a matter source to them. Assuming constant dilaton and imposing the section condition with respect to the regular coordinates leads to a universe dominated by momentum modes while imposing the section condition with respect to the dual coordinates naturally leads to a universe dominated by momentum modes. We show that both regimes have asymptotic behaviours related by T-duality. This hints towards defining a duality between the clocks considered in each regime and interpreting winding modes as being radiation from the point of view of a Euclidean time. 


\end{abstract}

\maketitle


\section{Introduction}

The T-duality symmetry \cite{Kikkawa:1984cp} plays an important role in string theory. For example, for strings on a torus of radius $R$, the symmetry implies that the spectrum of string states is unchanged if $R \rightarrow 1/R$ (in string units) and string momentum modes are interchanged with string winding modes. This symmetry is obeyed by string interactions, and it is assumed to be a symmetry of non-perturbative string theory (see e.g. \cite{Polchinski:1998rr, Boehm:2002bm}). 

T-duality is a key ingredient of the {\it String Gas Cosmology} (SGC) proposal \cite{Brandenberger:1988aj} for early universe cosmology. SGC in the present form (see e.g. \cite{Brandenberger:2008nx, Battefeld:2005av, Brandenberger:2011et} for reviews) is based on ideas coming from string thermodynamics. As is well known \cite{Hagedorn:1965st}, there is a maximal temperature for a gas of closed strings in thermal equilibrium, the {\it Hagedorn temperature} $T_H$. If we consider a box of strings of radius $R$, then the temperature $T(R)$ of a gas of strings in this box remains close to $T_H$ for a range of values of $R$ about $R =1$ (in string units), the range increasing as the entropy of the string gas increases. Hence, it was postulated in \cite{Brandenberger:1988aj} that the early phase of the universe might be a quasi-static hot string gas phase (see also \cite{Kripfganz:1987rh} for similar ideas). As was realized in \cite{Nayeri:2005ck}, thermal fluctuations in the quasi-static phase evolve into an approximately scale-invariant spectrum of cosmological fluctuations with a slight red tilt, and \cite{Brandenberger:2006xi} a scale-invariant spectrum of gravitational waves with a slight blue tilt. In this framework, the key size and shape moduli of extra dimensions can be stabilized in a natural way \cite{Watson:2003gf, Patil:2005fi, Brandenberger:2005bd}. What is missing in string gas cosmology, however, is a dynamical understanding of how the space-time background evolves. Our work is motivated by the aim to make progress on this issue. 

As already pointed out in \cite{Brandenberger:1988aj}, in string theory on a compact space there are two coordinates for each dimension of the topological space, firstly the coordinate $x$ associated with the momentum modes, and secondly the dual coordinate ${\tilde{x}}$ associated with the winding modes (note that for point particle theories there is only the coordinate $x$ associated with the momentum modes). It is hence to be expected that the dynamics of the background geometry of SGC should live in a doubled space including both $x$ and ${\tilde{x}}$ coordinates.

Double Field Theory (DFT) \cite{Siegel:1993xq, Siegel:1993th, Hull:2009mi} is an interesting proposal for a field theory living in doubled space. DFT is given (see e.g. \cite{Aldazabal:2013sca} for a review) by an action for a generalized metric in $2D$ space-time dimensions which is constructed from the metric, antisymmetric tensor field and dilaton (the ``background'') of the massless sector of a $D$ space-time dimensional string theory. In particular, after imposing a {\it section condition} the dynamical equations for the background reduce to those of supergravity (we will here focusing on bosonic supergravity).

In this paper we will couple a cosmological (i.e. homogeneous and isotropic) background of DFT to matter described by some energy density $\rho$ and some pressure $p$. We will study various ways of imposing a section condition, assuming that the dilaton is fixed. If we impose the section condition with respect to the regular coordinate (i.e. we assume that the variables do not depend on the dual coordinates), then we find that the equation of state of matter has to be that of regular radiation. On the other hand, if we impose the section condition with respect to the dual coordinates (i.e. we assume that the variables do not depend on the regular coordinates), then we find that the equation of state of matter has to be an equation of state dominated by winding modes. However, even though the equations of state differ, the background dynamics is the same, corresponding to a radiation-like expansion. 

We then speculate about a dynamical transition between the two branches of solutions, a correct description of which would have to go beyond the strict framework of DFT and would have to involve more stringy considerations. We argue that this transition might involve a complexification of the scale factor (see also \cite{Feldbrugge:2017kzv} where complexifications of the scale factor have been recently proposed to resolve the cosmological singularity).

The supergravity (SUGRA) equations of motion have been studied extensively for homogeneous and isotropic space-times (see e.g. \cite{Tseytlin:1991xk, Gasperini:2002bn}). If the antisymmetric tensor vanishes they reduce to the ones of dilaton gravity. Recently, we \cite{Brandenberger:2018xwl} have considered the coupling of dilaton gravity to a perfect fluid matter source whose equation of state is that expected from a gas of fundamental closed strings. This equation of state has the property that it is dominated by momentum modes at large values of the cosmological scale factor (equation of state $w = 1/d$, where $d$ is the dimension of space), by winding modes for small values (equation of state $w = - 1/d$), and has zero presssure for intermediate values. The resulting solutions were shown to be nonsingular in both the string and Einstein frames, when interpreted in terms of a dual time variable in the winding mode regime \footnote{In a related paper \cite{Brandenberger:2017umf} we showed that from this point of view point particle geodesics in DFT can be extended arbitrarily far into the past and future, indicating that DFT cosmology is geodesically complete.}. In this paper we go one step further and consider the cosmological equations in the context of double field theory.

DFT equations of motion for cosmology have been studied in \cite{Wu:2013sha}, but in the absence of matter sources. Matter sources have been explicitly included in the general DFT equations recently derived in \cite{Angus:2018mep}, where care was taken to have both background and matter terms considered in the DFT-invariant way. However, in that paper no cosmological solutions were considered. It is such solutions which we consider here. 

A word on notation: small Latin letters $i, j, ...$ indicate indices which run over the regular $D = d + 1$ space-time dimensions, capital letters $M, N, ...$ stand for indices which run over both the regular and the dual space-time dimensions. The cosmological scale factor is denoted by $a(t)$, where $t$ is the physical time. The dual time is denoted by ${\tilde{t}}$. The equation of state parameter $w$ is $w = p / \rho$, where $\rho$ and $p$ are energy density and pressure, respectively.

\section{Short review of Double Field Theory}

The DFT action unifies the metric $g_{ij}$, the two-form
$b_{ij}$ and the dilaton $\phi$ by rewriting these fields in an $O(D,D,\mathbb{R})$ covariant way, where $D$ is the number of spacetime dimensions, and it reduces to the supergravity action
if there is no dependence on the dual coordinates\footnote{This is usually done by imposing the section condition. See \cite{Angus:2018mep} for a derivation of the section condition from a special class of translation invariance allowed by the $O(D,D)$ symmetry group.}. It is
given by,
\begin{eqnarray}\label{DFT_action}
S \, = \,  \int dxd\tilde{x} e^{-2d}\mathcal{R},
\end{eqnarray}
where $d$ contains both the dilaton $\phi$ and the determinant of the metric,
\begin{equation}
e^{-2d} \, = \, \sqrt{-g} e^{-2 \phi} \, ,
\end{equation}
and where \cite{Hohm:2010jy},
\begin{eqnarray}
\mathcal{R} & = & \frac{1}{8}\mathcal{H}^{MN}\partial_{M}\mathcal{H}^{KL}\partial_{N}\mathcal{H}_{KL}-\frac{1}{2}\mathcal{H}^{MN}\partial_{M}\mathcal{H}^{KL}\partial_{K}\mathcal{H}_{NL}\nonumber \\
 &+&4\mathcal{H}^{MN}\partial_{M}\partial_{N}d-\partial_{M}\partial_{N}\mathcal{H}^{MN}-4\mathcal{H}^{MN}\partial_{M}d\partial_{N}d\nonumber \\
 &+&4\partial_{M}\mathcal{H}^{MN}\partial_{N}d,
\end{eqnarray}
where the generalized metric, $\mathcal{H}_{MN}$, is defined as,
\begin{eqnarray}
\mathcal{H}_{MN} \, = \,
\begin{bmatrix}
g^{ij} & -g^{ik}b_{kj} \\
b_{ik}g^{kj} & g_{ij}-b_{ik}g^{kl}b_{lj} \\
\end{bmatrix} \, ,
\end{eqnarray}
having its $O(D,D)$ index structure being lifted or lowered by the matrix $\eta^{MN}$, defined as,
\begin{eqnarray}
\eta^{MN} \, = \,
\begin{bmatrix}
0 & \delta^{\;\;j}_i \\
\delta^{i}_{\;\;j} & 0\\
\end{bmatrix} \, .
\end{eqnarray}

Throughout the rest of this paper we will be working with double spacetime coordinates.


\section{Dual Cosmology}

The cosmological background equations of motion coming from DFT are the same as the ones coming from SUGRA for the dynamical fields in terms of the momentum coordinates. Thus, all the solutions found in that context can be automatically brought into the DFT framework. What differs in the latter is that the underlying geometry is  (2d+1)-dimensional and given by the following line element, 
\begin{eqnarray}
dS^{2}&=&-dt^{2}+\mathcal{H}_{MN}dX^{M}dX^{N}
\nonumber \\
&=&-dt^{2}+a^{2}\left(t\right)d\vec{x}^{2}+a^{-2}\left(t\right)d\tilde{x}^{2}.
\end{eqnarray}
We already considered some physical aspects of this metric in previous work. In \cite{Brandenberger:2018xwl} we showed that the cosmological solutions of supergravity in the presence of a perfect fluid matter gas with the equation of state appropriate to SGC \footnote{This equation of state corresponds to radiation with $w = 1/d$ for large values of $a(t)$ and to a gas of winding modes with $w = - 1/d$ in the small $a$ limit.} are nonsingular in the string frame, and can be given a nonsingular interpretation in the Einstein frame making use of a dual time ${\tilde{t}}$ which replaces the regular time $t$ for small values of the scale factor. Making use of this dual time, it was shown earlier \cite{Brandenberger:2017umf} that point particle geodesics can be extended infinitely far both into the future and towards the past. 

If we start with two time coordinates, the cosmological ansatz for the line element is
\begin{eqnarray}
dS^{2}&=&\mathcal{H}_{MN}dX^{M}dX^{N}
\nonumber\\
&=&-dt^{2} - d\tilde{t}^2+a^{2}(t,\tilde{t})d\vec{x}^{2}+a^{-2}(t,\tilde{t})d\tilde{x}^{2},
\end{eqnarray}
where now the generalized metric is also defined in terms of the temporal component of the space-time metric. 

The vacuum equations of motion for DFT in the presence of a dual-time associated to the winding sector are given by  \cite{Wu:2013sha}, 
\begin{align}
4d^{''}-4d^{'2}-(D-1)\tilde{H}^{2}+4\ddot{d}-4\dot{d}^{2}-\left(D-1\right)H^{2} & =0 \nonumber\\
\left(D-1\right)\tilde{H}^{2}-2d^{''}-(D-1)H^{2}+2\ddot{d} & =0 \nonumber\\
\tilde{H}^{'}-2\tilde{H}d^{'}+\dot{H}-2H\dot{d} & =0 \label{DFTVacuum},
\end{align}
where a prime denotes a derivative with respect to $\tilde{t}$, and a dot a
derivative with respect to $t$. Also, note that $\tilde{H}\equiv a'/a$ and $2d \equiv 2\phi - (D-1)\ln{a}$ is the shifted dilaton. To derive these equations, we need to vary the DFT action (\ref{DFT_action}) with respect
to $d$, $g_{tt}$ and $g_{ii}$, respectively, assuming a cosmological background while taking into account the constraint,
\begin{equation}
\mathcal{H}_{MN}\mathcal{H}^{NP}=\delta_{M}^{P} \label{GenMetricConstraint},
\end{equation}
which in our case simply implies $g_{\mu\nu}g^{\nu\rho}=\delta_{\mu}^{\rho}$\footnote{Note that associated to $d\tilde{t}$ is $g_{\tilde{t}\tilde{t}}\equiv g_{tt}^{-1}.$}. 

Comparing (\ref{DFTVacuum}) to standard string cosmology equations \cite{Tseytlin:1991xk} in the presence of matter, we propose coupling the above equations to matter by the following prescription, 
\begin{align}
& 4d^{''}-4d^{'2}-(D-1)\tilde{H}^{2}+4\ddot{d}-4\dot{d}^{2}-\left(D-1\right)H^{2}  =0 \nonumber\\
& \left(D-1\right)\tilde{H}^{2}-2d^{''}-(D-1)H^{2}+2\ddot{d}  =\frac{1}{2}e^{2d}E(t,\tilde{t}) \nonumber\\
& \tilde{H}^{'}-2\tilde{H}d^{'}+\dot{H}-2H\dot{d}  =\frac{1}{2}e^{2d}P(t,\tilde{t}), \label{DFTMatter}
\end{align}
where $P$ and $E$ are the pressure and energy associated to the matter sector, respectively\footnote{In \cite{Tseytlin:1991xk} matter was introduced via the matter action  $S=\int dt\sqrt{-g_{tt}}F\left(\log a,\beta\sqrt{-g_{tt}}\right),$
where $F$ is the one loop free energy. In principle, this prescription can be extended in the presence of double time (see appendix \ref{appendix1}). It is left for a future work \cite{DFTCosmology}
to consider its full covariant formulation.}. Note that they are also function of $\tilde{t}$. One might wonder
if these quantities would have their dual counterparts. That is not the case, since we know that $g_{tt}=g_{\tilde{t}\tilde{t}}^{-1}$
and $g_{ii}=g_{\tilde{i}\tilde{i}}^{-1}$ due to (\ref{GenMetricConstraint}). Therefore, it is easy to see that,
\begin{equation}
\frac{\delta}{\delta g_{\tilde{t}\tilde{t}}}=-g_{tt}^{2}\frac{\delta}{\delta g_{tt}},
\end{equation}
so that varying (\ref{DFT_action}) with respect to $g_{\tilde{t}\tilde{t}}$ would result in the same equation of motion derived by varying it with respect to $g_{tt}$. The same follows for the other components\footnote{Regardless, we can still think of the dual counterparts as a matter of definition,
even though they are not independent of the normal ones. In particular,
it is easy to see that $\tilde{\rho}=-\rho$, after assuming $g_{tt}=-1$,
and $\tilde{p}=-a^{4}\left(t,\tilde{t}\right)p,$ given $g_{ii}=a^{2}\left(t,\tilde{t}\right).$}\footnote{Note that fully covariant equations coupling matter
to DFT background were proposed in \cite{Angus:2018mep}. They
arise from varying a generalized action consisting
of geometrical action plus matter action with respect
to the generalized metric. It reduces to a generalization of supergravity when the section condition is imposed in the supergravity frame, where the gravitational charge can also be acquired through couplings with the dilaton and the antisymmetric tensor field. The cosmology of such framework will be discussed at \cite{DFTCosmology}.  In comparison, our equations,
derived for homogeneous and isotropic space-times, are not fully covariant, as it is further discussed and explored in \cite{Bernardo:2019pnq}. In order to define the energy density and pressure we need to make explicit use of the frame which was being considered. Hence, the
equations of \cite{Angus:2018mep} have a larger scope of application.}. 

Having the equations of motion in double space-time,
we need now to consider the imposition of the section condition. Typically
it is imposed in the so called supergravity frame, which means assuming that the dynamical fields do not depend on the dual coordinates. However, one could equally well
consider the imposition of any $O\left(D,D\right)$ rotation of the
section condition, in particular assuming that there is no dependence on the regular coordinates rather than no dependence on the dual coordinates. We will consider
the resulting dynamics in each frame and draw an interpretation in terms of
the overall existence of either momentum or winding modes in the next section. 

Before we proceed, a few comments are in order here. There are two different takes we can consider regarding DFT: as a fundamental theory or as a mathematical framework that connects T-dual solutions of string theory. If one assumes DFT is fundamental, then changing between frames is the same as considering different gauge choices, and even though the solutions look different, this is just due to a gauge choice \cite{Park:2013mpa}. On the other hand, one can consider DFT as a theory that connects distinct physical solutions, which in the context of DFT would be accounted for in different frames. For instance, it is known that there are string backgrounds which are non-geometric and do not have a good supergravity description, yet they are captured by alternative frame choices \cite{Hull:2009sg,Kachru:2002sk,Rudolph}. In fact, an explicit example of such interpretation was considered in \cite{Berman:2014hna}, where there is a natural choice of frame picked by the string/wave solution as one approaches its core, being the dual frame the natural one. We will consider the latter approach throughout the rest of this paper. 

In order to simplify our reasoning, we will be considering the dilaton to be already stabilized. Thus, we have $\text{2\ensuremath{\dot{d}}}=-\left(D-1\right)H$
and $2d^{'}=-\left(D-1\right)\tilde{H}$, and  (\ref{DFTMatter}) become, 
\begin{align}
& 2\left(\tilde{H}^{'}+\dot{H}\right)+D\left(\tilde{H}^{2}+H^{2}\right)  =0 \nonumber\\
& \left(\tilde{H}^{2}-H^{2}\right)+\left(\tilde{H}^{'}-\dot{H}\right)  =\frac{1}{2\left(D-1\right)}G\rho\left(t,\tilde{t}\right) \nonumber\\
& \left(\tilde{H}^{'}+\dot{H}\right)+\left(D-1\right)\left(\tilde{H}^{2}+H^{2}\right)  =\frac{G}{2}p\left(t,\tilde{t}\right),\label{DFTMatterFixDil}
\end{align}
where $G$ depends on $\phi=\phi_{0}$, the fixed value of the dilaton. The most important feature to be noticed in these equations is the asymmetry between the regular and dual coordinates dependence in the second equation. Now we consider each particular
frame. 

\subsection{Supergravity frame: large radius limit}

The mass spectrum of a closed string in a one-dimensional space, compactified on a circle, is 
\begin{equation}
    M^2 = (N + \tilde{N} -2) + p^2 \frac{l_s^2}{R^2} + w^2 \frac{l_s^2}{\tilde{R}^2}\label{MassSpec},
\end{equation}
where $N, \tilde{N}$ correspond to oscillatory modes of the string, $p$ corresponds to its momentum modes, associated to the center of mass motion, and $w$ corresponds to winding modes, accounting for the number of times the string has wrap itself around the compact dimension in a topologically non-trivial way.

We expect that as the scale factor becomes larger\footnote{Small and large here are always in relation to the string length $l_s$.} only
momentum modes will be energetically favorable (considering
that the radius of the compact dimensions would also become larger), as it can be easily seen from (\ref{MassSpec}).  In this case, we hope that only the $t$-dependence should be relevant,
given the $\tilde{t}$ was introduced exactly to tackle the winding
modes dynamics from a T-dual perspective. This is typically called
supergravity frame, but here we are putting forward an interpretation
associated with this frame. We will do similarly in the next sub-section when considering
the winding-frame. 

After imposing the section condition on the $\tilde{t}$-coordinates, the equations of motion (\ref{DFTMatterFixDil}) reduce to the standard string cosmology equations for a stabilized dilaton \cite{Tseytlin:1991xk},
\begin{align}
2\dot{H}+DH^{2} & =0 \nonumber\\
-H^{2}-\dot{H} & =\frac{G}{2\left(D-1\right)}\rho(t)\nonumber\\
\dot{H}+\left(D-1\right)H^{2} & =\frac{\text{G}}{2}p(t), 
\end{align}
which imply the following equation of state, 
\begin{equation}
w=\frac{1}{D-1},\label{RadEOS}
\end{equation}
corresponding to a radiation-like universe. This leads to the scale factor evolving as
\begin{equation}
a\left(t\right) \varpropto t^{2/D}.
\end{equation}
Evidently, the continuity equation is,
\begin{equation}
\dot{\rho}+DH\rho=0,
\end{equation}
and the energy density redshifts as radiation,
\begin{equation}
\rho\left(a\right)\varpropto a^{-D}(t).
\end{equation}

The above result is not surprising since it is well known from studies of dilaton-gravity that an expanding universe the dilaton can only be constant if the equation of state of matter is that of radiation.

\subsection{Winding-frame: small radius limit}

Now we consider what we call the winding-frame, in which we impose
the section condition on the regular coordinates. We expect this frame to be a good description for the regime in which winding modes
dominate, corresponding to the limit of small radius as seen in (\ref{MassSpec}). In this case the equations of motion become, 
\begin{align}
2\tilde{H}^{'}+D\tilde{H}^{2} & =0 \nonumber\\
\tilde{H}^{2}+\tilde{H}^{'} & =\frac{1}{2\left(D-1\right)}G\rho(\tilde{t}) \nonumber\\
\tilde{H}^{'}+\left(D-1\right)\tilde{H}^{2} & =\frac{1}{2}Gp(\tilde{t}), 
\end{align}
implying the following equation of state,
\begin{equation}
w=-\frac{1}{D-1}, \label{WinEOS}
\end{equation}
which corresponds to a fluid composed only of winding modes. We thus see that constant dilaton in the winding frame is only consistent if the equation of state of matter is that of a gas of winding modes. This is quite surprising, since this resulted from assuming a regime in which the $t$-dependence is gone, not an \emph{a priori} assumption about the matter content, reinforcing the interpretation of this frame being associated to a dynamics ruled by only winding modes. 

Due to the asymmetry between the frames seen in (\ref{DFTMatterFixDil}), and having an equation of state given by winding modes, which is the negative of what we have for radiation, the continuity equation reads, 
\begin{equation}
\rho'+D\tilde{H}\rho=0,
\end{equation}
which implies that the energy density will also redshifts as radiation, 
\begin{equation}
\rho (a)\varpropto a^{-D}(\tilde{t}),
\end{equation}
despite corresponding to winding modes. This differs from what we find in usual cosmology, where (\ref{WinEOS}) implies $\rho\varpropto a^{2-D}$ instead. 

Note that our result explicitly shows that the universe is T-dual, \textit{i.e.}, a universe characterized by a winding equation of state in the dual coordinates behaves exactly the same as a universe dominated by momentum modes in the regular coordinates. However, it is important to realize that as we take the limit of small scale factor in the momentum frame, which approaches a singularity, the winding frame expands due to the scale factor duality  $a(t) \rightarrow a^{-1}(\tilde{t})$ \cite{Buscher:1987sk,Buscher:1987qj}. Imposing the section condition separates both solutions, but if matter in the universe is made of both winding and momentum modes, a smooth transition, not possible in standard DFT, is needed. We investigate this further in a future work \cite{PhysicalClockPaper}.

In order to solve for the scale factor, first we notice that unlike the momentum case, the corresponding Friedmann-like equation has
a minus sign, 
\begin{equation}
\tilde{H}^{2}=-\frac{G}{\left(D-2\right)\left(D-1\right)}\rho. \label{WindFriedEq}
\end{equation}
Thus, we see that either $\tilde{H}$ is complex and $\rho>0$ (which
implies $p<0$, as usual for winding modes) or $\tilde{H}$ is real
but $\rho<0$ (and $p>0$). 

Considering $\tilde{H}$ to be complex, then we can work with the following
ansatz,
\begin{equation}
a\left(\tilde{t}\right)=\tilde{A}\left(\tilde{t}\right)e^{i\theta\left(\tilde{t}\right)},
\end{equation}
so that the Friedmann-like equation becomes,
\begin{align}
\tilde{H}_{\tilde{A}}^{2}-\theta^{'2} & =-g\rho_{0}\tilde{A}^{-D}\cos\left(D\theta\right) \nonumber\\
2\tilde{H}_{\tilde{A}}\theta^{'} & =g\rho_{0}\tilde{A}^{-D}\sin\left(D\theta\right), \label{WindFriedEq}
\end{align}
where $g\equiv G/\left(D-2\right)\left(D-1\right).$ Note that for
$\theta=\pi/D,$ the second equation vanishes identically and the
first equation becomes,
\begin{equation}
\tilde{H}_{\tilde{A}}^{2}=g\rho_{0}\tilde{A}^{-D},
\end{equation}
so that,
\begin{equation}
a\left(\tilde{t}\right)=\tilde{a}_{0}\tilde{t}^{2/D}e^{i\pi/D},
\end{equation}
where $\tilde{a}_{0}$ is a constant in this regime. 

Let us take a moment to analyze what we have just derived. Before,
for the momentum case, we obtained the solution,
\begin{equation}
a_{m}\left(t\right)=a_{0}t^{2/D}e^{i\theta_{m}},
\end{equation}
where $\theta_{m}=0,$ since the solution was real. Writing
both solutions together, we have,
\begin{equation}
\begin{cases}
a_{m}\left(t\right) & =a_{0}t^{2/D}\\
a_{w}\left(\tilde{t}\right) & =\tilde{a}_{0}\tilde{t}^{2/D}e^{i\pi/D}.
\end{cases}
\end{equation}
Now, remembering that the scale factor solution associated to the
winding modes is the reciprocal of the one associated to the momentum ones (and
ignoring those arbitrary constants for the moment),
\begin{align*}
a_{m} & \rightarrow a_{w}^{-1},
\end{align*}
and we conclude that the solutions are dual given, 
\begin{equation}
t\rightarrow\tilde{t}^{-1}e^{-i\pi/2}.
\end{equation}
Therefore, quite surprisingly, we can also interpret that the winding
scale factor solution corresponds to a Wick rotation of the reciprocal
of the momentum time-coordinate. Since $\theta$ is actually a dynamical variable
for us, this rotation happens dynamically as it will be shown below.

For the general case, we still need to solve (\ref{WindFriedEq}). Combining the equations, we see that 
\begin{equation}
\tilde{H}_{A}^{2}=\frac{g\rho_{0}\tilde{A}^{-D}}{2}\left[1-\cos\left(D\theta\right)\right].
\end{equation}
The solution for $\theta$ is given by,
\begin{equation}
\theta\left(\tilde{t}\right)=\pm\frac{2}{D}\arccos\left[\left(\frac{\tilde{A}}{\tilde{A}_{0}}\right)^{-D/2}\right],
\end{equation}
where $\tilde{A_{0}}$ is a constant. Therefore, 
\begin{equation}
\tilde{H}_{A}^{2}=g\rho_{0}\tilde{A}^{-D}\left[1-\left(\frac{\tilde{A}}{\tilde{A}_{0}}\right)^{-D}\right],
\end{equation}
which implies that,
\begin{equation}
\tilde{A}\left(\tilde{t}\right)=\left[\tilde{A}_{0}^{D}-\frac{1}{4}D^{2}C_{1}^{2}+\frac{D^{2}}{4}g\rho_{0}\tilde{t}^{2}\pm i\frac{D^{2}C_{1}}{2}\sqrt{g\rho_{0}}\tilde{t}\right]^{1/D},
\end{equation}
for some arbitrary constant $C_{1}$. In particular, given that we
have chosen $\tilde{A}\left(\tilde{t}\right)$ to be real, this constant
should be set to $0$ since it would not be there in the first place if
we had complied with the assumption that $\tilde{H}_{A}$ was real. We see that for
large $\tilde{t}$ we recover the typical radiation solution as expected
with a complex phase.

Having this general solution is quite helpful also to understand the
particular case we considered above, the one $\theta=\pi/D$. In principle,
we would like to have this happenning dynamically as opposed to just
fixing $\theta$ by hand. To see that,
let us take the large $\tilde{t}$ limit, then
\begin{equation}
\tilde{A}\left(\tilde{t}\right)\rightarrow\tilde{t}^{2/D},
\end{equation}
which implies that 
\begin{equation}
\theta\left(\tilde{t}\right)\rightarrow\pm\frac{2}{D}\arccos\left(\frac{1}{\tilde{t}}\right)\xrightarrow[\tilde{t}\rightarrow\infty]{}\pm\frac{\pi}{D}.
\end{equation}
Therefore, this shows that it is the case, indeed, that deep in the
winding regime this phase is singled out and the oscillations in the
scale factor cease to exist. Also, this shows that our temporal duality
defined above appears due to the dynamics of our solutions.

Finally, as an illustration, we could have considered $\rho<0$ in (\ref{WindFriedEq}), ending up
with,
\begin{equation}
a\left(\tilde{t}\right)=\tilde{a}_{0}\tilde{t}^{2/D}.
\end{equation}
For this case, the winding and radiation solutions would be dual under the following identification,
\begin{equation}
\tilde{t}\rightarrow\frac{1}{t}.
\end{equation}
This further motivates the heuristic
arguments considered in \cite{Brandenberger:2017umf,Brandenberger:2018xwl}. We see that for positive energy density, the temporal parameter space is complex, while for negative energy density it is $\mathbb{R}^2$.

\section{Comments}

Considering the results from the last section, we can speculate about a different interpretation of these findings. It has been argued before in the context of quantum cosmology and quantum gravity that the ground state of the wave function of the universe should correspond to Euclidean geometry \cite{Hartle:1983ai}, which would have a non-zero probability of tunneling to a de Sitter state of continual expansion. 

In fact, such proposals motivated studies about classical change of the metric's signature \cite{Ellis:1991sp}. In particular, it has been observed that the equation of state for a perfect fluid gets a minus sign when the underlying geometry is Euclidean. Therefore, the radiation equation of state in Euclidean space would mimic the equation of state of winding modes. 

If we run this reasoning backwards, we could speculate that winding modes should be understood as radiation when time is Euclidean. The transition we observe from winding frame to momentum frame would correspond simply to a change of the signature of the metric. This is further developed in \cite{PhysicalClockPaper}.

\section{Conclusion}

Double Field Theory can be interpreted as a natural generator of T-dual solutions once the section condition is imposed in one or another set of coordinates. We investigated these different solutions after considering a cosmological ansatz for the metric. We have shown that in both frames we have a radiation-like dynamics, but with different equations of state. On one side, the universe is dominated by winding modes while in the other one, by momentum modes. This is exactly what one would expect. The small scale factor limit in one frame approaches a singularity while in the dual frame it expands to an infinite volume. The T-dual mapping between the two frames provides further evidence for the connection between the two time coordinates pointed out previously in \cite{Brandenberger:2017umf, Brandenberger:2018xwl}.  

\section*{Acknowledgement}
\noindent

This research is supported by the IRC - South Africa - Canada Research 
Chairs Mobility Initiative Grant No. 109684. The research at McGill is 
also supported in part by funds from NSERC and from the Canada Research 
Chair program. GF acknowledges financial support from CNPq (Science 
Without Borders) and PBEEE/Quebec Merit Scholarship. GF also wishes to 
thank the University of Cape Town where most of this work was developed. 
R.C. acknowledges financial support by the SARChI NRF grantholder. A. W. 
gratefully acknowledges financial support from the Department of 
Science and Technology and South African Research Chairs Initiative of 
the NRF.
\\


\appendix 

\section{A note about T-dualizing matter} \noindent\label{appendix1}

Typically, matter is introduced in SUGRA by the following action \cite{Tseytlin:1991xk},
\begin{equation}
S=\int d^{D}x\sqrt{-g}e^{-2\phi}f=\int dt\sqrt{-g_{tt}}F\left(\log a,\beta\sqrt{-g_{tt}},\phi\right),
\end{equation}
where $F$ is the free energy,
\[
F=\int d^{D-1}xa^{D-1}e^{-2\phi}f.
\]

Formally, we can think of a T-dual covariant generalization of it by defining
the following action, 
\begin{equation}
S=\int d^Dx d^D\tilde{x} e^{-2d}\mathcal{F},
\end{equation}
with $\mathcal{F}$ depending on both sets of coordinates. Then, we can write
\begin{align}
S & =\int dxd\tilde{x}dtd\tilde{t}\sqrt{g}e^{-2\phi}\mathcal{F}\nonumber \\
 & =\int dt\sqrt{-g_{tt}}\left(\int d^{D-1}\tilde{x}d\tilde{t}F\right)
\end{align}
or 
\begin{equation}
S=\int d\tilde{t}\sqrt{-\frac{1}{g_{tt}}}\left(\int d^{D-1}\tilde{x}dtF\right).
\end{equation}
where $F$ is also function of both sets of coordinates. 

Finally, the standard definitions of the energy and pressure of the system follow as usual,
\begin{eqnarray}
E(t,\tilde{t}) & = & -2 \frac{\delta F}{\delta g_{00}}
\\
P_i(t, \tilde{t}) & = & - \frac{\delta S}{\delta \ln a_i}.
\end{eqnarray}

\bibliographystyle{unsrt}

\bibliography{References}

\end{document}